\documentclass[aps,prc,twocolumn,showpacs,showkeys,preprintnumbers,amsmath,amssymb,a4paper]{revtex4}
\usepackage{graphicx}% Include figure files
\usepackage{bm}% bold math
\usepackage{epsfig}

\newcommand{\be}{\begin{eqnarray}}
\newcommand{\ee}{\end{eqnarray}}

\def\bfg #1{{\mbox{\boldmath $#1$}}}

\begin{document}
\title{ Elastic $\bar p d$ scattering and total $\bar p d$ cross sections
}

\author{Yu.N.~Uzikov$^{1,2}$ and J.~Haidenbauer$^{3,4}$}

\affiliation{
$^1$Laboratory of Nuclear Problems, Joint Institute for Nuclear
Research, 141980 Dubna, Russia\\
$^2$Physics Department, Moscow State University, 119991 Moscow, Russia \\
$^3$Institute for Advanced Simulation, Forschungszentrum J\"ulich, 
D-52425 J\"ulich, Germany \\
$^4$Institut f\"ur Kernphysik and J\"ulich Center for Hadron Physics, Forschungszentrum J\"ulich,
D-52425 J\"ulich, Germany \\
}
\date{\today}

\begin{abstract}
Elastic $\bar p d$ scattering is studied within the Glauber
theory based on the single- and double $\bar p N$ scattering mechanisms. 
The full spin dependence of the elementary $\bar p N$ scattering amplitudes is
taken into account and both the $S$- and $D$-wave components of the deuteron 
are considered. 
The treatment of the spin dependence is done in a (properly modified) formalism 
developed recently by Platonova and Kukulin for the $pd\to pd$ scattering process.
Predictions for differential cross sections and the spin observables 
$A_y^d$, $A_y^{\bar p}$, $A_{xx}$, $A_{yy}$ are presented for antiproton beam
energies between 50 and 300 MeV, using amplitudes generated from the $\bar N N$ 
interaction model developed by the J\"ulich group. 
Total polarized cross sections are calculated utilizing the optical theorem.
The efficiency of the polarization buildup for antiprotons in a storage ring is
investigated.
\end{abstract}


\maketitle
  
\section{Introduction}

  The present investigation is motivated by the plans of the PAX collaboration \cite{barone} 
to measure the transversity of the proton (antiproton) in double-polarized Drell-Yang processes
 at an upgrade of the FAIR facility in Darmstadt.
  In order to achieve this aim an intense polarized beam of antiprotons is required.
  A possibility to overcome the experimental challenge to obtain such a polarized beam 
  is seen in scattering of antiprotons off a polarized $^1$H target in rings \cite{frank05}.
  Analogous experiments performed for the proton case by the FILTEX collaboration \cite{FILTEX}
  at 23 MeV and a recent COSY study where protons were scattering off a polarized 
  hydrogen at 49 MeV \cite{frankpc} showed that indeed a polarized (proton) beam can 
  be achieved via the so-called spin-filtering effect, i.e. by exploiting the fact 
  that via the scattering process protons are removed (lost) from the ring at different 
  rates for different initial polarization states \cite{FILTEX}. 
  According to theoretical interpretations \cite{MS}
 of the data \cite{FILTEX,frankpc}, the polarization buildup effect appears
 solely due to the hadronic interaction of the incoming proton with the target.

Whereas the spin dependence of the nucleon-nucleon ($NN$) interaction is very well known
at the considered energies, that allows one to calculate reliably the
spin-filtering effect for protons,
there is practically no corresponding information for 
the antiproton-nucleon ($\bar NN$) interaction. For this reason a test experiment 
for the spin-filtering effect in the antiproton-hydrogen interaction is planned 
at the AD ring at the CERN facility \cite{AD1,AD}.

In view of the unknown spin dependence of the $\bar pN$ interaction, the 
interaction of antiprotons with a polarized deuteron is also
of interest for the issue of the antiproton polarization buildup. 
This option was discussed in a previous paper by us \cite{ujh2009}.
In that work the single-scattering approximation was used
for the calculation of the polarized total $\bar p d$ cross sections for 
energies in the region 50--300 MeV. The spin dependence of the
elementary $\bar p N$ amplitudes was taken into account in this approximation
only in collinear kinematics using the $\bar N N$ interaction model of the J\"ulich
group \cite{Hippchen,Mull1,Mull2,Haidenbauer2011}. The $\bar p N$ double-scattering 
effects were only accounted for in the computation of the {\it unpolarized} total and differential 
cross sections and found to be in the order of 10-15\% \cite{ujh2009}. 
Spin observables for $\bar p d$ elastic scattering
and shadowing effects (double scattering) in polarized total cross sections  
were not considered in that work. A calculation of such observables, including
double-scattering effects, is the main aim of the present work. 
Indeed corresponding results are certainly desirable, specifically 
in view of the prospect that a discrimination of existing models 
of the $\bar p N$ interaction could be feasible on the basis of a comparison 
with expected data \cite{AD}. 
Antiproton polarization buildup in the context of $\bar p d$ scattering was 
also studied in a recent work by Salnikov~\cite{salnikov}, utilizing the
results from the Nijmegen $\bar pp$ partial wave analysis \cite{Timmermans}
from 1994. (Note that an updated partial wave analysis has been presented
recently by Zhou and Timmermans \cite{Timmermans2}). 

In the present paper we consider elastic $\bar p d$ scattering
within the Glauber theory of multi-step scattering 
\cite{Glauber1,Sitenko1,GlauberFranco}, 
taking into account the full spin-dependence of the elementary $\bar p N$ scattering amplitudes.
There are several studies of the accuracy of the Glauber
theory in the literature \cite{harrington,osborn,lakkolyb73,bianconi95,fonseca07,overmeire07} 
which demonstrate that corrections to the eikonal approximation, which is the basis
of this theory, are small in the region of intermediate energies about $\sim  1$ GeV. 
The reliability of the Glauber approach at intermediate energies was studied 
recently in Ref.~\cite{elster2008} via a comparison with rigorous Faddeev calculations 
for the case of identical spinless bosons interacting by means of a simple
Malfliet-Tjon interaction potential. The results of Ref.~\cite{elster2008} for
such a ``bosonic'' $nd$ system show that even at rather low energies, $100 \sim 200$ MeV, the
difference between the Faddeev and the Glauber calculations is just about 10-15\% 
for the total cross section. Rather good agreement 
was found also for the differential cross section in the forward hemisphere,
excluding the region of the diffraction minimum \cite{elster2008}.
 
For antiproton-nucleus scattering the Glauber theory can be applied
 at lower energies \cite{laksmbiz81,mahalanabis} as compared to the proton-nucleus reaction.
The amplitude for elastic scattering of antiprotons off nuclei is 
strongly peaked in forward direction due to strong annihilation effects
 in the $ \bar p N$ interaction, supporting the applicability 
 of the eikonal approximation (see Ref.~\cite{ujh2009} and reference therein).  
In the present work the $S$- and $D$-wave components of the deuteron
and both the single- and double $\bar p N$ scattering mechanisms are taken into account.
The treatment of the spin dependence is based on a proper modification of the
formalism developed recently by Platonova and Kukulin \cite{pkuk,pkuk2}
for $pd$ elastic scattering.  
In their papers the formalism was succesfully applied for describing
spin observables of the $pd\to pd$ process at 250--1000 MeV.
An independent confirmation of the findings of \cite{pkuk} 
was reported recently by us \cite{prague12}.
  
The spin dependence of the $\bar pd\to \bar p d$
 amplitude is very similar to that
 for $ pd \to  p d$ scattering, except for the contribution of the
 charge-exchange channel $\bar p p \leftrightarrow \bar n n$ which,
 however, can be taken into account straightforwardly.
 In the present work we consider the differential cross section 
and the spin observables $A_y^d$, $A_y^{\bar p}$, $A_{xx}$, $A_{yy}$ for the 
${\bar p}d\to {\bar p}d$ process. Those observables are 
evaluated here for antiproton beam energies from 50 to 300 MeV 
employing $\bar p N$ amplitudes generated from the J\"ulich $\bar N N$ model
\cite{Mull2}. 
The total polarized cross sections $\sigma_1$, $\sigma_2$, and  $\sigma_3$ 
are calculated on the basis 
of the optical theorem. We also investigate the efficiency of the polarization 
buildup for antiprotons in a storage ring. Here Coulomb effects are 
taken into account within the formalism described in Ref. \cite{ujh2009}. 

The paper is structured in the following way:
In Sect. II we introduce briefly the used formalism. In particular, we
point out the differences that occur between its application to the
$p d$- and to the $\bar p d$ systems.
In Sect. III results for $\bar p d$ scattering are presented.
First we discuss the issue of the applicability of the Glauber
theory. Specifically, we assess the angular range for which differential
observables can be reliably calculated within this approach. 
Then predictions for the differential cross section
and the spin observables $A_y^d$, $A_y^{\bar p}$, $A_{xx}$, $A_{yy}$ 
are given based on elementary $\bar p N$ amplitudes taken from the
J\"ulich $\bar NN$ models A and D.
Finally, our results for the total $\bar p d$ cross section are
provided, including those for the polarized case. We also provide predictions 
for the polarization degree of the antiproton beam which is the decisive quantity 
for the spin-filtering method. The paper ends with a short Summary. 
Relations between amplitudes and considered observables are given in an Appendix.
    
\section{Elements of the formalism for $\bar pd \to \bar p d$ scattering}

\begin{figure}
%\hspace{0.5cm}
%\vspace {0.5cm} 
\mbox{
\epsfig{figure=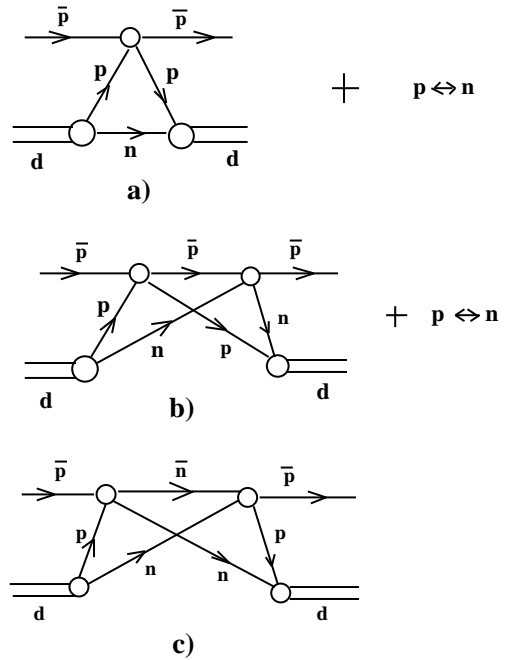,width=0.75\linewidth,angle=0}
}
\caption{Mechanisms for $\bar p d$ elastic scattering:
 single scattering (a), double scattering (b), and  
charge-exchange (c).}
\label{mechs}
\end{figure}

For the $\bar p d\to \bar pd$ process we use the formalism developed in Ref. \cite{pkuk}
for the process $pd \to  p d$, taking into account the specific differences that
arise for the $\bar p d$ collision.
Within the Glauber theory \cite{GlauberFranco} the scattering matrix 
for elastic $\bar p d$ scattering is given by the following matrix element  
\begin{equation}
\label{glafi}
M_{\bar p d}({\bf q})= \langle f|\hat M({\bf q}, {\bf s})|i \rangle,
\end{equation}
evaluated between definite initial $|i\rangle$ and final $|f\rangle$ states of
the $\bar p d$ system. Here
${\bf q}$ is the momentum transferred from the initial to the final deuteron
in the $\bar pd$ collision and ${\bf s}$ is the impact parameter of the projectile.
In Eq.~(\ref{glafi}) and in the following we suppress the dependence of $M_{\bar p d}$ 
on the total energy to simplify the notation. 
The transition operator $\hat M $ can be written as
\begin{eqnarray}
\label{Ffi}
&&{\hat M}({\bf q}, {\bf s})=\nonumber \\
&& \exp{(\frac{1}{2}i{\bf q}\cdot{\bf s})}M_{\bar p p}({\bf q})+
\exp{(-\frac{1}{2}i{\bf q}\cdot{\bf s})}M_{\bar p n}({\bf q})+\nonumber \\
&&\frac{i}{2\pi ^{3/2}}\int \exp{(i{\bf q}'\cdot{\bf s})}
\Bigl [M_{\bar p p}({\bf q}_1) M_{\bar p n}({\bf q}_2)+\nonumber \\
&&M_{\bar p n}({\bf q}_1) M_{\bar p p}({\bf q}_2)-
M_{\bar p p\to \bar n n}({\bf q}_1) M_{\bar n n\to \bar p p}({\bf q}_2) \Bigr ]
d^2{\bf q}'. \nonumber \\ 
\end{eqnarray}
Here $M_{\bar p N}({\bf q})$ ($N=p, n$) is the 
$\bar p N$ scattering matrix,  $M_{\bar p p\to \bar n n}$ ($M_{\bar n n\to \bar p p}$)
is the scattering matrix of
the charge-exchange process ${\bar p p\to \bar n n}$ ($\bar n n\to \bar p p$),
and ${\bf q}_1={\bf q}/2-{\bf q}'$ and ${\bf q}_2={\bf q}/2+{\bf q}'$ are
the transferred momenta in the first and second
$\bar p N$ collision, respectively, in the double-scattering terms. The first two 
plane-wave terms correspond to the single-scattering mechanism, while the three 
terms in the integral represent the double-scattering mechanism. 
These transitions correspond to the diagrams depicted in Fig. \ref{mechs}.
It is assumed in Eq.~(\ref{Ffi}) that the deuteron wave function in the matrix element
of this operator does not contain the isospin part explicitly. Therefore
the last term in Eq.~(\ref{Ffi}) appears with a negative sign coming from the
product of the isospin factors at the $dpn$ vertices.

The elementary scattering matrix for elastic $\bar p N$ scattering is 
given by the following expression:
\begin{eqnarray}
 M_{\bar p N}&=&A_N+(C_N\bfg \sigma_1+C_N'\bfg \sigma_2)\cdot {\bf \hat n}\nonumber \\
&+&B_N(\bfg \sigma_1\cdot {\bf \hat k})(\bfg \sigma_2\cdot {\bf \hat k}) \nonumber \\
&+&(G_N-H_N)(\bfg \sigma_1\cdot {\bf \hat n})(\bfg \sigma_2\cdot {\bf \hat n})\nonumber \\
&+&(G_N+H_N)(\bfg \sigma_1\cdot {\bf \hat q})(\bfg \sigma_2\cdot {\bf \hat q}) \ .
\label{pbarN}
\end{eqnarray}
Here $\bfg \sigma_1$ ($\bfg \sigma_2$) is the Pauli matrix acting on the spin of the
$\bar p$ ($N$) states, $N=p,n$. The unit vectors are defined by 
${\bf \hat k}=({\bf k}_i+{\bf k}_f)/|{\bf k}_i+{\bf k}_f|$,
${\bf \hat q}=({\bf k}_i-{\bf k}_f)/|{\bf k}_i-{\bf k}_f|$,
and ${\bf \hat n}=[{\bf \hat k}\times {\bf \hat q}]$, where ${\bf k}_i$ (${\bf k}_f$) 
denotes the momentum of the incident (outgoing) antiproton.
The one for charge-exchange, $M_{\bar pp\leftrightarrow \bar nn}$, has the same spin 
structure as given in Eq.~(\ref{pbarN}). 
We denote the corresponding amplitudes on the right-hand side of Eq.~(\ref{pbarN}) 
by $A_c$, $C_c$, etc. in the following for simplicity reasons. 

The amplitudes in Eqs.~(\ref{glafi})--(\ref{pbarN}) are normalized as in 
Ref. \cite{pkuk} and related to the corresponding differential cross sections via 
\begin{eqnarray}
\label{diffpd}
\frac{d\sigma}{dt}=\frac{1}{6} tr M_{\bar p d}M_{\bar p d}^+, \\
\label{diffpN}
\frac{d\sigma}{dt}=\frac{1}{4} tr M_{\bar p N}M_{\bar p N}^+, 
\end{eqnarray}
 where $t=-{\bf q}^2$ is the squared four-momentum transfer in the $\bar p d$ or
$\bar p N$ systems, respectively. 
The factor in front of the integral in Eq.~(\ref{Ffi}) differs from that 
in the original Glauber theory \cite{GlauberFranco} due to a different normalization
of the invariant scattering matrices $M$ used in Eqs.~(\ref{glafi})--(\ref{diffpN}).
 
In general, there are 36 transitions in the $\bar pd \to \bar p d$ process
for the different spin states. When accounting for rotational invariance 
and invariance under parity- and time-reversal transformations,
the number of transition matrix elements for the $\bar pd \to \bar p d$ process 
reduces to twelve independent complex amplitudes $A_i$ ($i=1,\dots,12$), which can be 
introduced in the same way as for the $pd \to  p d$ process
\cite{ujh2009,pkuk} (see also the Appendix).
The amplitudes $A_i$ 
can be written in the form 
\begin{eqnarray}
A_i(q)&=& A^{(s)}_i(q) \nonumber \\
&+&\frac{i}{2\pi^{3/2}}\int d^2q'\Bigl (A_i^{(d)}({\bf q}, {\bf q}')
-A_i^{(c)}({\bf q},{\bf q }') \Bigr).
\label{ai}
\end{eqnarray} 
Here we introduced the scalar amplitudes for
single scattering (Fig.~\ref{mechs}a), $A^{(s)}_i(q)$, 
double scattering (Fig.~\ref{mechs}b), $A_i^{(d)}({\bf q}, {\bf q}')$, 
and double-scattering involving charge exchange (Fig.~\ref{mechs}c), 
$A_i^{(c)}({\bf q}, {\bf q}')$, using the same notations as in Ref.~\cite{pkuk}.

The amplitudes $A_i^{(s,d,c)}$ can be expressed via the elastic form
factors of the deuteron corresponding to the transitions 
$S\to S$, $S\to D$, $D\to S$, $D\to D$, from the initial to the final
deuteron state, and the elementary amplitudes of $\bar pN$ scattering. 
The explicit expressions can be found in
Ref.~\cite{pkuk} in the appendices A and B and, therefore, we refrain
from reproducing those lengthy formulae here. 
Specifically, the formulae for the single-scattering mechanism
$A_i^{(s)}$ (Fig.~\ref{mechs}a) are the same as
those for the $pd \to p d$ processs given in Ref. \cite{pkuk}
in Table I with the replacement of 
the elementary amplitudes of $pN$ elastic scattering by the 
corresponding ones for $\bar pN$.
However, the integral in Eq.~(\ref{ai}), accounting for the double-scattering 
mechanisms differs from that in Eq.~(A.1)
in Ref.~\cite{pkuk} by two aspects. First, the term $A_i^{(d)}$ in 
Ref.~\cite{pkuk} has an additional factor 2. 
Second, the permutations $n \leftrightarrow p$ are taken into account here
for $A^{(s)}_i(q)$ and $A_i^{(d)}({\bf q}, {\bf q}')$, as in Ref. \cite{pkuk},
but they do not apply for $A_i^{(c)}({\bf q}, {\bf q}')$.
%5555555555%%%%%%%%%%%%%%%%%%%%%%%%%%%%%%%%%%%%%%%%%%%%%%%%
%%%%%%%%%%%%%%%%%%%%%%%%%%%%%%%%%%%%%%%%%%%%%%%%%5

In order to explain the reason for these two differences between the 
$\bar pd \to \bar p d$  and  $pd \to p d$ amplitudes,
let us discuss first the $pd\to pd$ process following Ref.~\cite{pkuk}.   
In the $pd\to pd$ process there are three double-scattering amplitudes, 
which we denote symbolically as
\begin{eqnarray}
M_{pp}({\bf q}_1)\times M_{pn}({\bf q}_2)&+&  
M_{pn}({\bf q}_1)\times M_{pp}({\bf q}_2)  \nonumber \\ 
 -M_{pn\to np}({\bf q}_1)&\times& M_{np\to pn}({\bf q}_2) \ . 
\label{treepd}
\end{eqnarray}

The small-angle charge-exchange scattering matrix $M_{pn\to np}$ can be
rewritten in terms of the small-angle scattering matrices 
$M_{pp\to pp}\equiv M_{pp}$ and $M_{pn\to pn}\equiv M_{pn}$ in the form \cite{glauberfranco67}
 $M_{pn\to np}({\bf q})=M_{np\to pn}({\bf q})=M_{pp}({\bf q})-M_{pn}({\bf q})$.
(The double-charge-exchange scattering matrix vanishes in the approximation
$M_{pp}({\bf q})=M_{pn}({\bf q})$,
i.e. if one disregards the isospin dependence of the $pN$ amplitude.)
Therefore, the third term in Eq.~(\ref{treepd}) can be written as
\begin{eqnarray}
M_{pn\to np}({\bf q}_1)&\times &
M_{np\to pn}({\bf q}_2)=\nonumber \\
M_{pp}({\bf q}_1)\times M_{pp}({\bf q}_2)&+&
M_{pn}({\bf q}_1)\times M_{pn}({\bf q}_2)\nonumber \\
-M_{pp}({\bf q}_1)\times M_{pn}({\bf q}_2)&-&M_{pn}({\bf q}_1)\times M_{pp}({\bf q}_2) \ .
\label{chepd}
\end{eqnarray}
The term $M_{pn\to np}({\bf q}_1)\times M_{np\to pn}({\bf q}_2) $  
enters the full $pd$ amplitude with opposite sign with respect to the
first two terms given in Eq.~(\ref{treepd}) due to the
additional permutation $p\leftrightarrow n$
in the deuteron vertex. 
Furthermore, the first and the second terms on the right hand side of 
Eq.~(\ref{chepd}) constitute the term called the "charge-exchange"
amplitude in Ref.~\cite{pkuk} and are denoted by $A_i^{(c)}$.
The third and the fourth terms from
Eq.~(\ref{chepd}) can be absorbed into the first and second 
terms in Eq.~(\ref{treepd}), respectively,
and this leads to the factor $2$ in front 
of the $A_i^{(d)}$ amplitude in \cite{pkuk}.

Coming back to the process $\bar p d \to \bar p d$, one should mention
that in this case there are also three double-scattering terms
in Eq. (\ref{Ffi}), namely 
\begin{eqnarray}
M_{\bar p p}({\bf q}_1)\times M_{\bar p n}({\bf q}_2), \ \  
M_{\bar p n}({\bf q}_1)\times M_{\bar p p}({\bf q}_2), \nonumber \\ 
M_{\bar pp\to \bar nn}({\bf q}_1)\times 
M_{\bar n n\to \bar p p}({\bf q}_2) \ .
\label{treeantpd}
\end{eqnarray}
The first two terms constitue
the amplitude $A^{(d)}_i$ in Eq.~(\ref{ai}). The detailed formulae for
$A^{(d)}_i$ ($i=1,\dots, 12$) coincide with those given in 
Table II of Ref.~\cite{pkuk} for the $pd\to pd$,
where each term ($A_N,\, B_N,\,C_N,\, C_N^\prime, \, G_N,\,H_N$)
from the $pN\to pN$ amplitude ($N=p,n$) has to be replaced by 
the corresponding term of the scattering matrix $M_{\bar pN\to \bar p N}$ 
given in Eq.~(\ref{pbarN}). 

The last term in Eq.~(\ref{treeantpd}) gives rise to the $A^{(c)}_i$ amplitude 
 in Eq.~(\ref{ai}) and, like for the $pd\to pd$ process,
 enters the full amplitude also with the opposite sign with respect 
 to the first two amplitudes in Eq.~(\ref{treeantpd}) for the same 
 reason as in the $pd\to pd$. 
 In the present calculation the term with the charge-exchange
 amplitudes $\bar p p \leftrightarrow \bar n n$ is not expressed through
 the small-angle scattering amplitude, as in the $pd\to pd$ case,
 but calculated straightforwardly
 and, therefore, does not lead to the factor $2$ in front of 
 $A^{(d)}_i$ in Eq.~(\ref{ai}) in contrast to Ref.~\cite{pkuk}. 
The formulae for $A^{(c)}_i$ ($i=1,\dots, 12$) are the same as those given in 
Table III of Ref.~\cite{pkuk}, where each product $A_n({\bf q}_2)A_n({\bf q}_1)$, etc.,
has to be replaced by the corresponding terms $A_c({\bf q}_2)A_c({\bf q}_1)$, etc., 
of the charge-exchange scattering matrix $M_{\bar pp\leftrightarrow \bar nn}$.
 Finally, we want to emphasize that contrary to Ref.~\cite{pkuk} the 
 total double-scattering term enters Eq.~(\ref{ai}) with positive sign.
 The relations of these amplitudes $A_i(q) $ with the spin observables considered 
 in the present paper are given in the Appendix.

%%%%%%%%%%%%%%%%%%%%%%%%%%%%%%%%%%%%%%%%%%%%%%%%%%%%%%%%%%%%%
\section{ Numerical results and discussion}

\subsection{General remarks and differential observables}

 Earlier studies \cite{laksmbiz81,mahalanabis,robson} and also our previous 
 calculations \cite {ujh2009,ujhprm} were all done within the spinless 
 approximation for the elementary ${\bar p N}$ amplitude $M_{\bar p N}$, i.e.
 they used only $A_N$ from Eq.~(\ref{pbarN}), parameterized in the Gaussian form,
 and they considered only the $S$-wave part of the wave function of the 
target nucleus. Those investigations suggested that the Glauber theory
 allowed one to explain the differential cross sections in the forward hemisphere and 
the total unpolarized cross section for the reactions $\bar p d$,  $\bar p^3{\rm He}$ and
 $\bar p ^4{\rm He}$ even at rather low energies, i.e. down to 20--50 MeV of the incident 
 antiproton.
 For ${\bar p}~^4{\rm He}$ elastic scattering the first and the second diffraction peaks
 were explained by these calculations \cite{ujhprm}. On the other hand, attempts 
 to describe the second peak observed in $\bar p d$ elastic scattering at $179$~MeV 
\cite{bruge} were not quite that successful as documented in some 
 papers \cite{mahalanabis,robson}.

 In the present calculation we include the $S$- and $D$-wave components of
 the deuteron wave function and 
 we keep the full spin dependence of the $\bar pN$ amplitude as given in
 Eq.~(\ref{pbarN}). With regard to the wave function we use here the one of 
the CD Bonn potential as parameterized in Ref.~\cite{Machleidt} and, for 
test calculations, also the one of the Paris potential~\cite{Paris}. 
The $\bar pN$ amplitude is taken from two models developed by the J\"ulich group, 
namely A(BOX) introduced in Ref.~\cite{Hippchen} and D described in Ref.~\cite{Mull2}. 
Results for the total and integrated elastic ($\bar p p$) and
charge-exchange ($\bar p p \to \bar nn$) cross sections and also
for angular dependent observables for both models can be found in
Refs.~\cite{Hippchen,Mull2} while specific spin-dependent observables 
are presented in \cite{Haidenbauer2011}.
Model A as well as D provide a very good overall reproduction of the low- and 
intermediate energy $\bar N N$ data as documented in those works. 
 
\begin{figure}[hbt]
\mbox{\epsfig{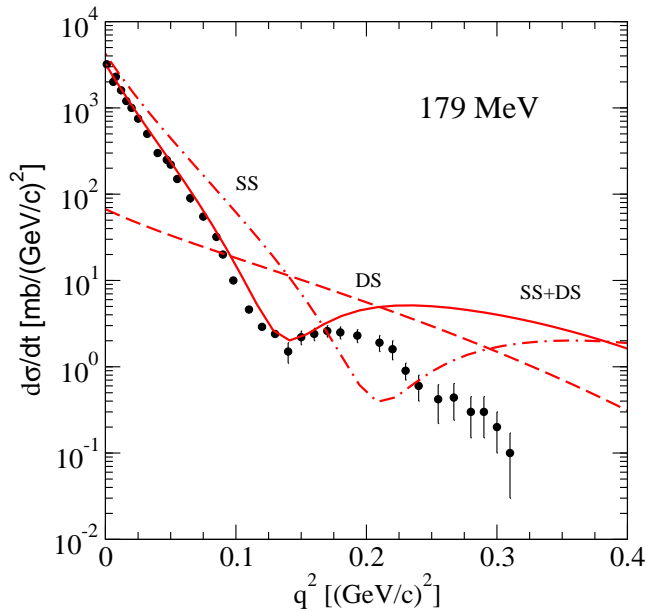}}
\caption{(Color online) Differential cross section of elastic ${\bar p} d$
scattering at 179~MeV versus the transferred momentum squared.
Results based on single-scattering (dash-dotted line), 
double-scattering (dashed) and the full (solid) 
Glauber mechanisms are shown. 
For the calculation the $\bar p N$ amplitude predicted 
by the J\"ulich model D is utilized. 
The data points are from Ref.~\cite{bruge}.
}
\label{pd179ns10}
\end{figure}

 An exemplary $\bar p d$ result demonstrating the role of the single-scattering (SS) and 
 double-scattering (DS) mechanisms is presented in Fig. \ref{pd179ns10}. 
 There, cross sections based on the SS- and the DS mechanisms and their 
 coherent sum (SS+DS) are displayed separately.
 One can see that the SS mechanism alone fails to explain the forward peak.
 However, the sum SS+DS describes it rather well. Obviously, 
 the DS mechanism, neglected in Ref.~\cite{ujh2009} in the calculation of the 
 spin-dependent total cross sections, has a sizable influence even in the 
 region of the forward peak.
 
%%%%55555555555555555555555555555555555555555555555555555

%%%%%%%%%%%%%%%%%%%%%%%%%%%%%%%%%%%%%%%%%%%%%
\begin{figure}
\vspace{0.7cm}
%\mbox{\epsfig{figure=ds179pnd.ps,width=0.99\linewidth,angle=0}
%\mbox{\epsfig{figure=s179pnk.eps,width=0.99\linewidth,angle=0}
\mbox{\epsfig{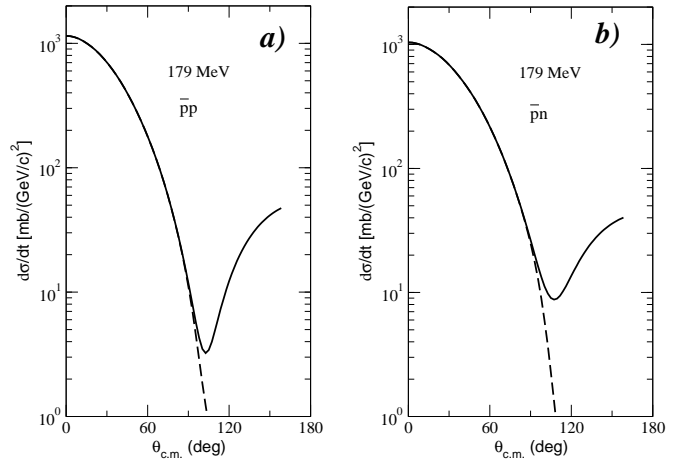}
}
\caption{Differential elastic ${\bar p} p$ (a) and ${\bar p} n$ (b)
cross sections at 179~MeV as predicted by model D (solid lines).
The dashed line shows the result with the smooth cutoff
with $q_0^2 = 0.21$~GeV/c and $\nu=10$,
corresponding to $\theta^{\bar p N}_{c.m.}= 105^\circ$,
 as explained in the text.}
\label{pn179}
\end{figure}

When accounting for the spin dependence of the $\bar p N$ interaction
one has to address also the reliability of the employed Glauber
approach, specifically, as far as the angular range is concerned. 
 In contrast to the spin-independent part of the amplitude Eq.~(\ref{pbarN}) 
 given by $A_N$ ($N=p,n$), most of the other amplitudes that give rise to the 
 spin dependence ($B_N$, $C_N$, $C_N^\prime$, $G_N$, $H_N$ in Eq.~(\ref{pbarN})) 
 do not exhibit
 a well-pronounced diffractive behaviour for antiproton beam energies 50--200 MeV,
 i.e. they do not decrease rapidly with increasing center-of-mass (c.m.) 
 scattering angle $\theta_{c.m.}$. 
 In fact, some of these amplitudes even increase with increasing $\theta_{c.m.}$
 and their magnitude is larger at $\theta_{c.m.}>90^\circ$ than at 
 $\theta_{c.m.}<90^\circ$. The typical behaviour of the differential ${\bar p} N$ 
 cross section can be seen in Fig.~\ref{pn179} for 179 MeV.
 This behaviour is basically the same for both J\"ulich models A and D and,
 therefore, we consider only model D in the following. 
 The visible rise of the cross section for backward angles is partly 
 due to those spin-dependent pieces of the ${\bar p} N$ amplitude and is 
 reflected in the corresponding ${\bar p} d$ results by the appearance of a 
 second, broad peak at large transferred momenta
 $q^2\sim 0.35$ (GeV/c)$^2$ coming from the SS contribution alone,
 as is seen in Fig.~\ref{pd179ns10}. 
 As a consequence, the second (``diffraction'') peak that appears in 
 the full calculation, including now the SS and DS mechanisms, originates 
 not only from the interference between the SS and DS amplitudes, as is
 usually the case for typical diffractive scattering,
 but is also related to that backward peak seen in the cross section
 of the (elementary) $\bar p N$ reaction. 
 We confirmed in test calculations within the SS mechanism, utilizing just 
 the amplitudes $A_N$ in Eq.~(\ref{pbarN}) 
 or a standard Gaussian-type representation of the total $\bar p N$ 
 amplitude in the spinless approximation,
% as given in Eq.~(\ref{gauss}),
 that then the differential cross sections do not demonstrate this behavior.

 One should note, that the Glauber diffraction theory of multi-step
 scattering is not suitable for taking into account backward scattering in 
 the elementary hadron-nucleon collision, because its basis is the eikonal 
 approximation. The deuteron elastic form factor suppresses 
 the contribution of the $\bar p N$ amplitudes at large angles, but at the 
 considered low and moderate energies this effect is not as strong as at 
 significantly higher energies for the same scattering angle.
 
%%%%%%%%%%%%%%%%%%%%%%%%%%%
\begin{figure}
\vspace{0.5cm}
%\hspace{1cm}
\mbox{\epsfig{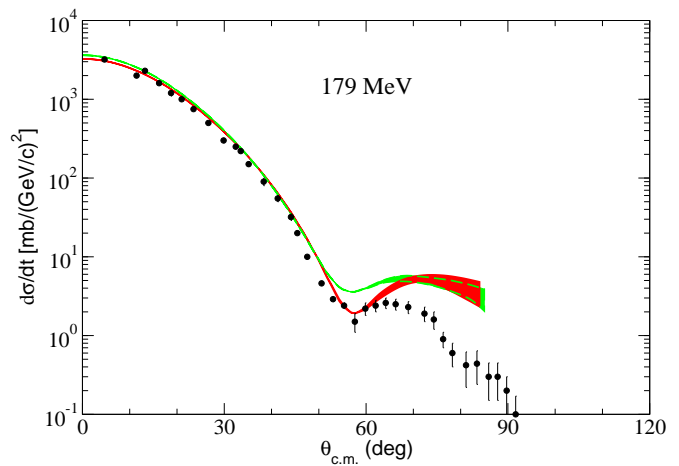}
}
\caption{(Color online) Differential cross section of elastic ${\bar p} d$
 scattering at 179~MeV versus the c.m. scattering angle.
 Results of our full calculation (including the SS+DS mechanisms)
 are shown based on the $\bar NN$ models model A (green/grey) and 
 D (red/black). The bands represent the sensitivy to variations
 of the large-angle tail of the $\bar p N$ amplitudes as discussed
 in the text. 
 The data points are taken from Ref. \cite{bruge}.}
\label{pd179q2}
\end{figure}
%%%%%%%%%%%%%%%%%%%%%%%%%%%%%%%%%%%%%%%%%%%%%%%%%%%%%%%%%%%%%555555

\begin{figure}
\vspace{0.5cm}
\mbox{\epsfig{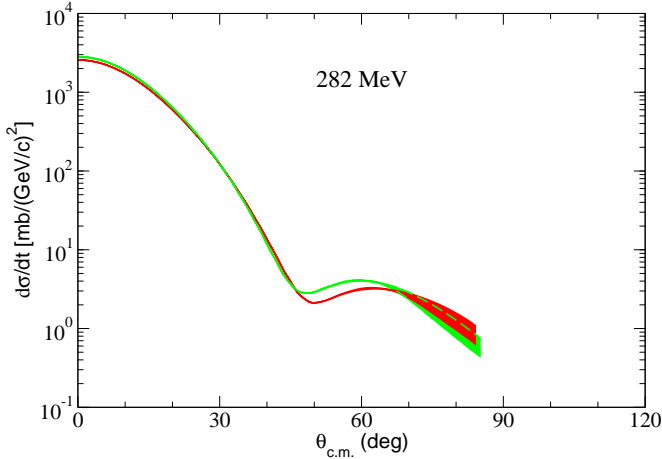}
}
\caption{(Color online) Differential cross section of elastic ${\bar p} d$
 scattering at 282~MeV versus the c.m. scattering angle.
 Same description of curves as in Fig. \ref{pd179q2}.}
\label{s282tet}
\end{figure}
%%%%%%%%%%%%%%%%%%%%%%%%%%%
\begin{figure}
\vspace{0.5cm}
\mbox{\epsfig{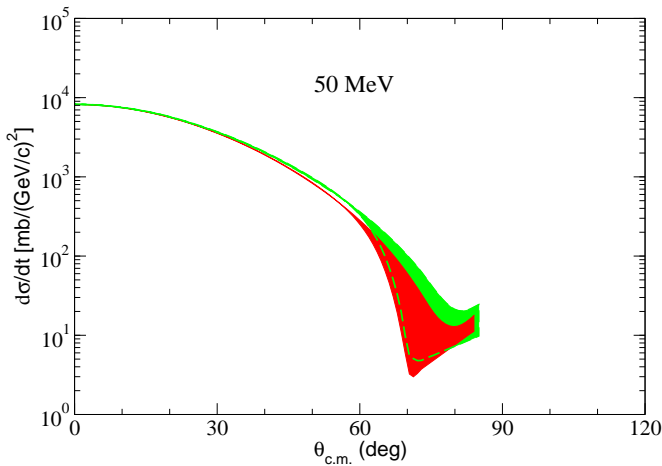}
}
\caption{(Color online) Differential cross section of elastic ${\bar p} d$
 scattering at 50~MeV versus the c.m. scattering angle.
 Same description of curves as in Fig. \ref{pd179q2}.}
\label{s050tet}
\end{figure}

The authors of Refs. \cite{pkuk,pkuk2} approximate all amplitudes in 
Eq.~(\ref{pbarN}) as a sum of Gaussians, e.g., 
\begin{eqnarray}
A_N(q) = \sum^n_{j=1} c_{A_N,j} \exp ( - d_{A_N,j}\, q^2), \ \ {\rm etc.} \ ,  
\label{Gauss}
\end{eqnarray}
where $c_{A_N,j}$ and $d_{A_N,j}$ are parameters fixed by a 
fit to the original $NN$ amplitudes. 
With such a series of Gaussians they are able to reproduce the original
amplitudes quite well in the forward hemisphere, i.e. up to angles of
$\theta_{c.m.}= 80^0-90^0$, as reported in \cite{pkuk2}. For larger angles
the differential $NN$ cross section based on those Gaussians drop off very
quickly, in accordings with the requirements by the Glauber model,
as shown in Fig. 4 of Ref.~\cite{pkuk2}. However, it remains unclear how
and, specifically, how fast the individual (parameterized) amplitudes drop off.
In particular, issues like the influence of the backward tail on the results 
or the related question of up to which angles one can trust the predictions
for $p d$ observables are not addressed in Ref.~\cite{pkuk}. 

%%%%%%%%%%%%%%%%%%%%%%%%%%%%%%%%%%%%%%%%%%%%%%%%%%%%%%%%%%%%%%%%%555555
\begin{figure*}
\mbox{\epsfig{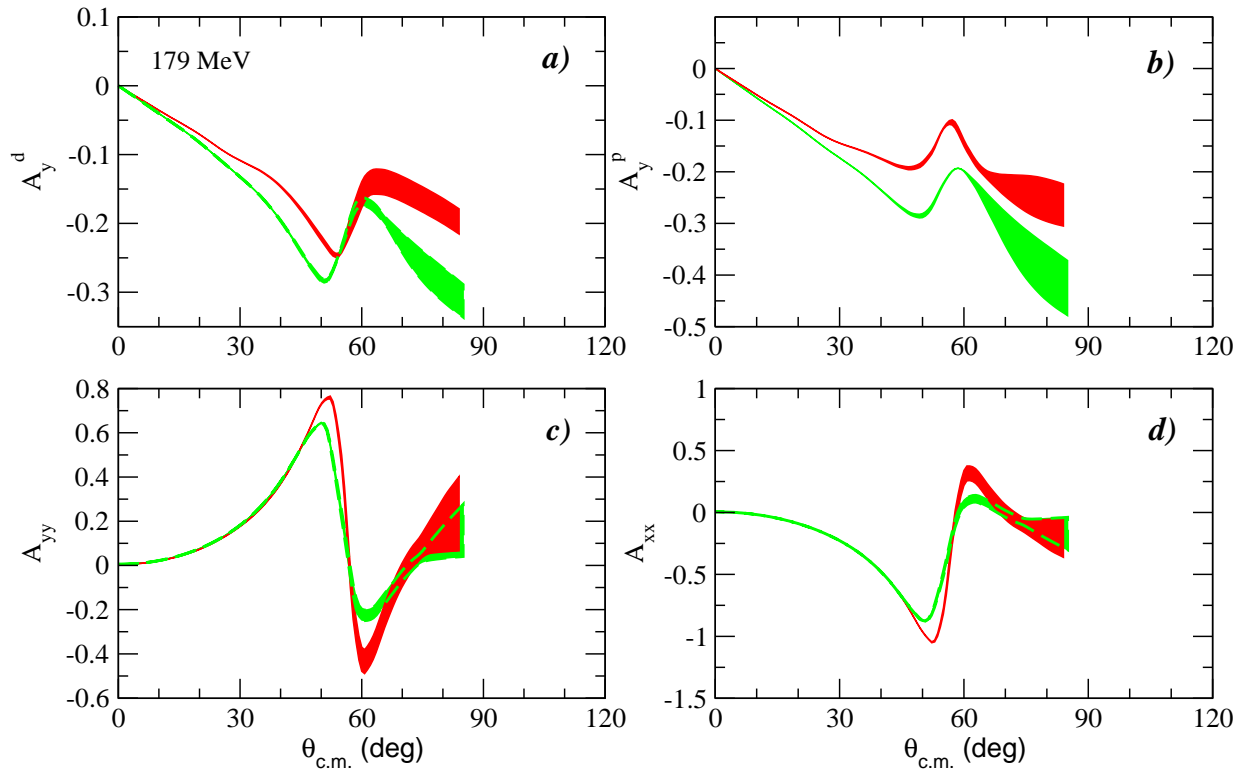}}
\caption{(Color online) Spin observables of elastic ${\bar p}d$ scattering at 179 MeV 
 versus the c.m. scattering angle:
$A_y^d$ (a), $A_y^{\bar p}$ (b), $A_{yy}$ (c), and $A_{xx}$ (d).
 Results of our full calculation (including the SS+DS mechanisms)
 are shown based on the $\bar NN$ models model A (green/grey) and
 D (red/black). The bands represent the sensitivy to variations
 of the large-angle tail of the $\bar p N$ amplitudes as discussed
 in the text.
 }
\label{ay179}
\end{figure*}

%%%%%%%%%%%%%%%%%%%%%%%%%%%%%%%%%%%%%%%%%%%%%%%%%%%%
\begin{figure*}[t]
\mbox{\epsfig{figure=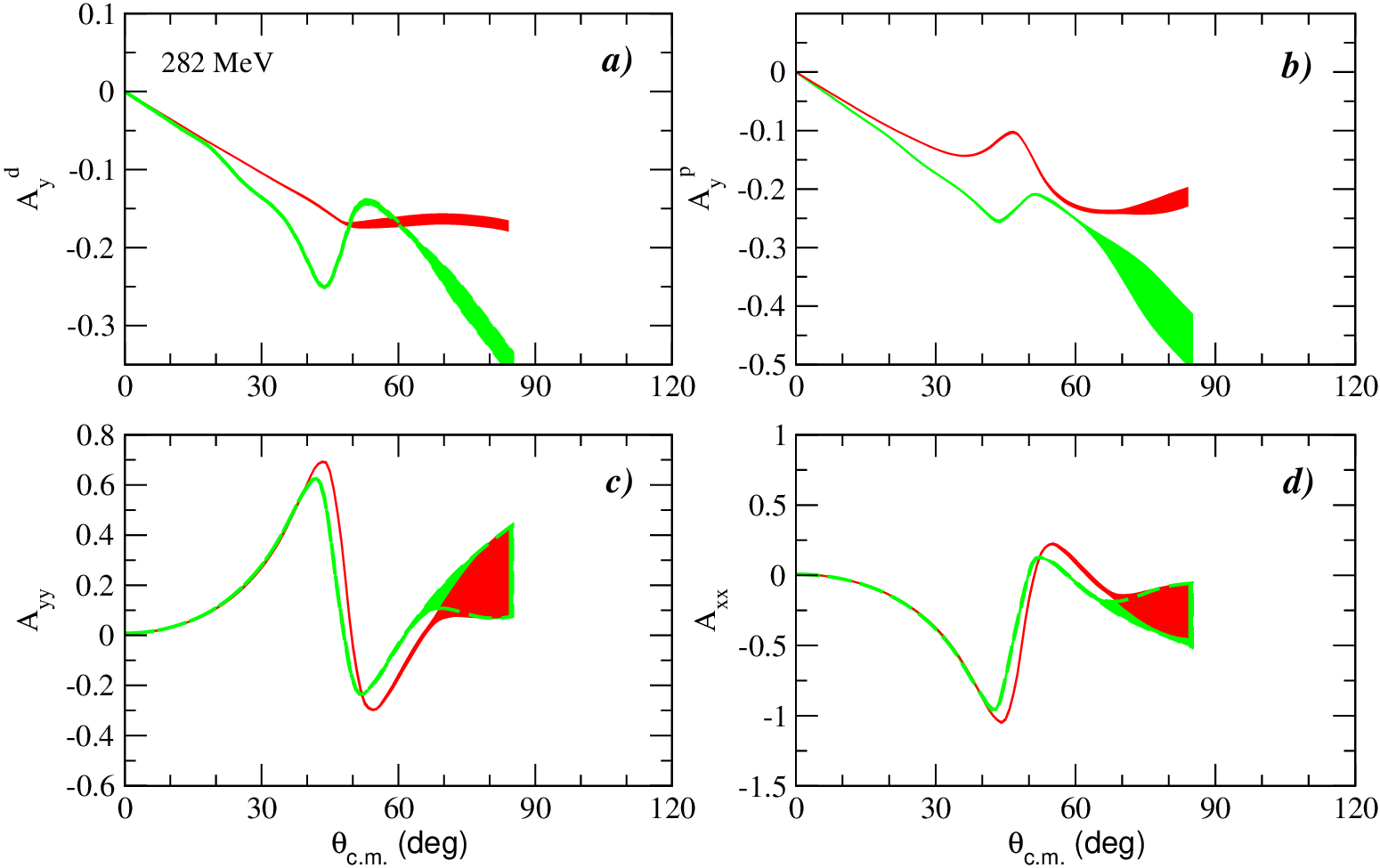,height=0.43\textheight, clip=}}
\caption{(Color online) Spin observables of elastic ${\bar p}d$ scattering at 282 MeV 
 versus the c.m. scattering angle.
 Same description of curves as in Fig. \ref{ay179}.
 }
\label{ay282}
\end{figure*}
%%%%%%%%%%%%%%%%%%%%%%%%%%%%%%%%%%%%%%%%%%%%%%%%%%%%
\begin{figure*}
\mbox{\epsfig{figure=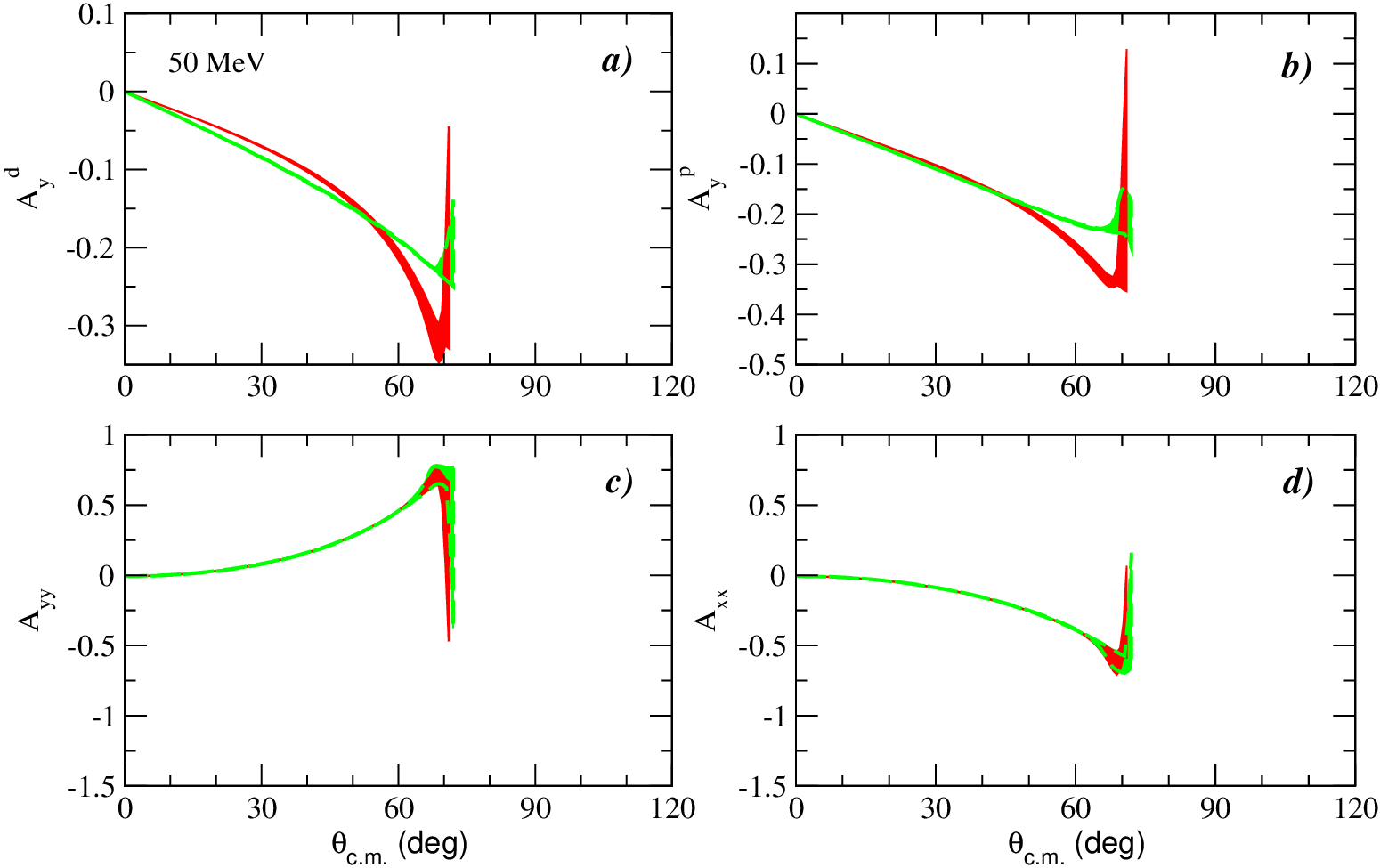,height=0.43\textheight, clip=}}
\caption{(Color online) Spin observables of elastic ${\bar p}d$ scattering at 50 MeV 
 versus the c.m. scattering angle.
 Same description of curves as in Fig. \ref{ay179}.
 }
\label{ay050}
\end{figure*}
%%%%%%%%%%%%%%%%%%%%%%%%%%%%%%%%%%%%%%%%%%%%%%%%%%%%
%
In the following we investigate this subject in the context of 
elastic $\bar p d$ scattering.
For this purpose we perform various calculations of this reaction 
within the Glauber theory in order to scrutinize the sensitivity 
of the considered $\bar p d$ observables to the large-angle region 
of the employed elementary $\bar p N$ amplitudes. 
Clearly, any such sensitivity is in contradiction with the assumptions
of the Glauber theory and tells us that the corresponding 
$\bar p d$ results are no longer reliable. 
We use also a Gaussian ansatz for representing the amplitudes
generated by the $\bar NN$ models A and D in analytical form. However, we aim 
at an excellent reproduction of the original amplitudes over the whole angular
range. This means that typically we have to use 10 or even 12 terms in the
sum in Eq.~(\ref{Gauss}), instead of $n=5$ which was taken in \cite{pkuk,pkuk2}.

We examine the sensitivity by varying the large-angle behavior of the
$\bar NN$ amplitudes that enter our Glauber calculation for $\bar p d$. 
In a first series of calculations a {\it smooth cutoff} on the angular
region is introduced by multiplying the employed $\bar p N$ amplitude 
by the factor $F(q)=1/[1+(q^2/q_0^2)^{\nu}]$. This is done in the evaluation of 
the SS mechanism as well as for the DS mechanism, i.e. in the 
two-dimensional integral in Eq.~(\ref{ai}). The lowest value for 
the cutoff momentum $q_0$ is chosen in such a way that it corresponds 
to a $\bar p N$ c.m. scattering angle which is close to the position of the minimum
of the elementary $\bar pN$ cross section. For the case of 179 MeV, shown in 
Fig.~\ref{pn179} (see dashed line), this amounts to $\theta_{c.m.}\approx 105^\circ$. 
This choice for $q_0$ makes sure that at least the $\bar p N$ cross section
produced by those modified $\bar p N$ amplitudes shows a clear diffractive behaviour.
We then increase the cutoff momentum $q_0$ until the corresponding
angle $\theta_{c.m.}$ reached 180 degrees. In the actual calculations the exponent 
$\nu$ was varied in the range $\nu=10 \div 15$.

As an alternative to variations of $q_0$ we performed also calculations
with a {\it sharp cutoff} at the maximum transferred momentum in the 
physical region $q=2k_{\bar pN}$, where $k_{\bar pN}$ is the $\bar pN$ c.m. momentum. 

Finally, we employed different Gaussian parameterizations, varying the number
of terms between 6 and 12, and without cutoff. Those parameterizations of the
$\bar pN$ amplitudes differ from the ones considered above in the angular range
$90^\circ \leq \theta_{c.m.} \leq 180^\circ$ but, more importantly, for 
larger transferred momenta $q$ outside of the physical range, i.e. 
$q > 2k_{\bar pN}$. Note that the
integration in Eq.~(\ref{ai}) requires the $\bar pN$ amplitudes at 
any (large) $q$, however, as already said above, the deuteron wave function
strongly suppresses contributions to the integral from that region. 

The result of our analysis is summarized in Figs. \ref{pd179q2}--\ref{ay050}.
The bands represent the variation of the calculated $\bar p d$ observables 
due to the cut-off procedures described above. 
We regard these bands as a sensible guideline for estimating the angular region
where the Glauber theory is able to provide solid results for
a specific observable and where this approach starts to fail. 
In particular, they indicate when contributions from large angles start to 
become significant. Since such contributions are in contradiction with the basic 
approximations underlying the Glauber model, any sizable influence from
them undoubtedly marks the breakdown of this approach.

The above considerations suggest that within the Glauber 
approach reliable predictions can be obtained for the differential cross 
section (Figs. \ref{pd179q2}--\ref{s050tet})
and also for the spin observables
$A_y^d$, $A_y^{\bar p}$, $A_{xx}$, and  $A_{yy}$ (Figs. \ref{ay282}, \ref{ay179}, 
\ref{ay050})
for c.m. scattering angles up to $50^\circ-60^\circ$ in the $\bar p d$ system.
Obviously, within this angular region there is practically no sensitivity to the 
$\bar p N$ amplitudes in the backward hemisphere, in accordance with the
requirements of the Glauber approach. 
As expected, for larger angles where such a sensitivity is observed, 
the width of the corresponding bands are smaller for higher energies (Fig. \ref{s282tet}) 
and larger at lower energies (Fig. \ref{s050tet}). This feature can be seen
in case of the differential cross sections and also 
for the spin observables $A_y$, $A_{yy}$, $A_{xx}$ presented 
in Figs. \ref{ay179}, \ref{ay282}, and \ref{ay050}.
 
A comparison of our calculation with experimental data is only meaningful 
for such ${\bar p d}$ scattering angles where the 
Glauber approach works well. 
According to our calculations this region 
includes the whole diffractive peak in the differential cross section 
$d\sigma/dt$ at forward angles, for energies from around 50 MeV upwards. 
This finding is important for the issue of the polarization buildup
of antiprotons, because it validates the application of 
the optical theorem for evaluating the total
polarized cross sections based on the obtained forward $\bar p d$ amplitude.  
 
With regard to the measured differential cross section at 179 MeV,
see Fig.~\ref{pd179q2}, our Glauber
calculation describes quite well the first diffractive peak - for $\bar p N$
amplitudes generated from model A as well as for those of model D. 
The first minimum in the differential cross section, 
located at $q^2\approx 0.12-0.13$ (GeV/c)$^2$
(i.e. $\theta_{c.m.}\approx 55^\circ$), and the onset of the second 
maximum is explained only by model D. The obvious strong disagreement with 
the data at larger transferred momenta, $q^2> 0.15$ (GeV/c)$^2$, 
corresponding to $\theta_{c.m.}>60^\circ$, lies already in the region where
the Glauber theory cannot be applicable anymore -- for the
case of the spinless approximation for the $\bar pN$ amplitudes 
as well as when the spin-dependent amplitudes are included --
and, therefore, no conclusions can be drawn. 
Note that at lower energies, specifically at 50 MeV, the first minimum 
lies outside of the region where the Glauber approach can be trusted.
In this context let us also mention that the results shown in Fig. \ref{pd179ns10} 
were obtained without any cutoff. 
  
Some remarks on the spin-dependent observables presented in 
Figs. \ref{ay179}, \ref{ay282}, and \ref{ay050}:
 The results obtained for the vector analyzing powers
 $A_y^{\bar p}$ and $A_y^d$ indicate a strong model dependence.
 In contrast, the tensor 
 analyzing powers $A_{xx}$ and $A_{yy}$ exhibit a very similar behaviour for 
 both models A and D. We found that the
 spin-independent amplitudes dominate the latter observables 
 and the inclusion of the spin-dependent amplitudes has only a minor influence.
 Thus, the results obtained here for $A_{xx}$ and $A_{yy}$ 
 seem to be quite robust up to scattering angles of $60^\circ-70^\circ$.
 When the spin-dependent terms of the elementary $\bar p N$ amplitude 
($B_N$, $ C_N$, $C_N'$, $G_N$, $H_N$)
 are excluded, then the vector analyzing powers $A_y^{\bar p}$ and$A_y^d$ vanish.
 
At 50 MeV the uncertainties in the considered spin-dependent observables 
increase dramatically for angles around 65$^\circ$, in accordance with the
strong variations that one sees in the differential cross section 
(Fig. \ref{s050tet}), and, therefore, we do not show those quantities 
beyond 70$^\circ$. 

We looked also at the influence of the $D$-wave component of the
deuteron on the obtained results. In the differential cross
section the contribution due to the $D$-wave is rather small in forward direction, 
but increases with increasing scattering angle. For example, at 179 MeV the 
contribution by the $D$-wave amounts to around $30$\% of the absolute 
value in the region of the first minimum.
The tensor analyzing powers $A_{xx}$ and $A_{yy}$ are considerably reduced 
(by one order of magnitude) when the $D$-wave is neglected. Actually, these
observables practically vanish if, in addition, the spin-dependent terms of the 
elementary $\bar p N$ amplitude are omitted. 
For observables that exhibit a larger sensitivy to the $D$-wave component we
performed also test calculations with the wave function of the Paris potential
which has a somewhat larger $D$-wave probability \cite{Paris}. It turned out
that the sensitivity to differences in wave functions is, in general, fairly small. 
Even in case of those tensor analyzing powers they amount to variations in the
order of 2-4\% only and they occur predominantly at the minima (maxima).

\subsection{Total spin-dependent cross sections}

The total $\bar p d$ cross section is defined by 
 \cite{ujh2009}
\begin{equation}
\label{totalspin}
\sigma_{tot}=\sigma_0+\sigma_1{\bf P}^{\bar p}\cdot {\bf P}^d+
 \sigma_2 ({\bf P}^{\bar p}\cdot {\bf \hat  k}) ({\bf P}^d\cdot {\bf \hat k})+
\sigma_3 P_{zz}, 
\end{equation}
where ${\bf \hat k}$ is the unit vector in the direction of the antiproton beam,
 ${\bf P}^{\bar p}$ ($ {\bf P}^d$) is the polarization vector of the antiproton (deuteron),
 and $P_{zz}$ is the tensor polarization of the deuteron ($OZ||{\bf \hat k}$).
 The total unpolarized cross section $\sigma_0$ and the spin-dependent cross sections
 $\sigma_i$ ($i=1,2,3$) are calculated using the generalized optical theorem as described 
 in Ref.~\cite{ujh2009}. Note, however, that erroneous expressions for the $\sigma_i$
have been given and used in that work (Eqs.~(19)-(20) in \cite{ujh2009}).
Specifically, the correct $\sigma_1$ and $\sigma_2$ which are shown in the present work
are of opposite sign to those given in Ref.~\cite{ujh2009}, see the Appendix for details
and for the correct expressions.
 
%sigma1.eps
%%%%%%%%%%%%%%%%%%%%%%%%%%%%%%%%%%%%%%%%%%%%%%%%%%%%%%%%%%%%%
\begin{figure}[hbt]
\mbox{\epsfig{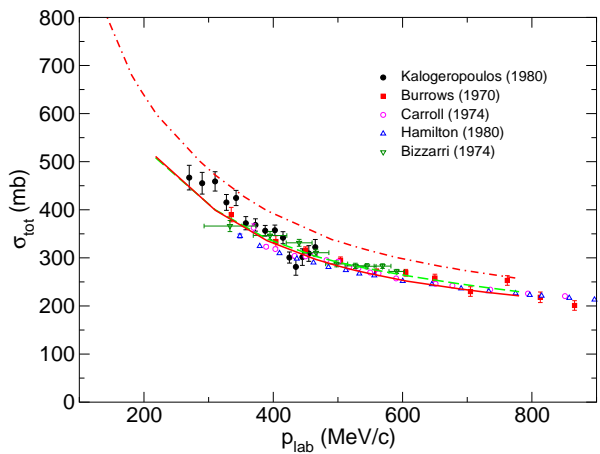}}
\caption{(Color online) Total unpolarized $\bar p d$ cross section versus the antiproton 
laboratory momentum.
Results of our full calculation (including the SS+DS mechanisms)
are shown based on the $\bar NN$ models A (dashed line) 
and D (solid line).
The results obtained for the SS mechanism alone based on
model D is indicated by the dash-dotted line. 
Data are taken from Refs.
\cite{BizzarriNC74,Kalogeropoulos,Burrows,Carroll,Hamilton}.
 }
\label{totsigm0}
\end{figure}
%%%%%%%%%%%%%%%%%%%%%%%%%%%%%%%%%%%%%%%%%%%%%%%%%%%%%%%%%%%%%
\begin{figure}[hbt]
\mbox{\epsfig{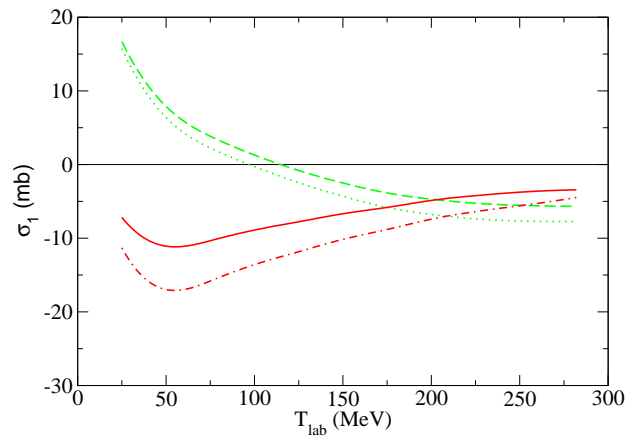}}
\caption{(Color online) Total $\bar p d$ cross section $\sigma_1$ versus the antiproton 
kinetic energy in the laboratory system. 
Results are shown based on the $\bar NN$ models A (green/grey) and D (red/black).
Calculations for the SS mechanism alone are indicated by dotted (A) and 
dash-dotted (D) lines while the full calculations (SS+DS mechanisms) are 
given by the dashed (A) and solid (D) lines. 
 }
\label{totsigm1}
\end{figure}
%%%%%%%%%%%%%%%%%%%%%%%%%%%%%%%%%%%%%%%%%%%%%%%%%%%%%%%%%%%%%

\begin{figure}[hbt]
\mbox{\epsfig{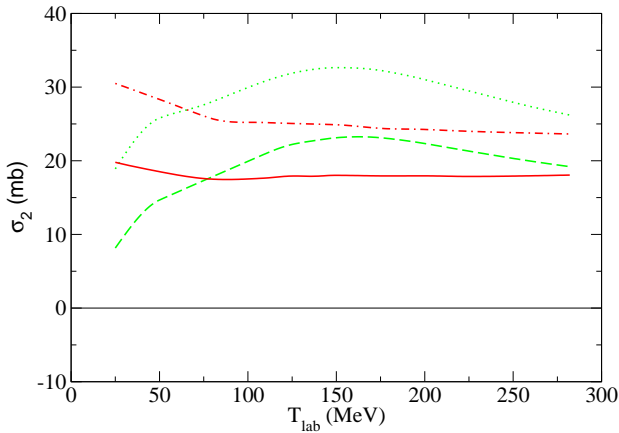}}
\caption{(Color online) Total $\bar p d$ cross section $\sigma_2$ versus the antiproton
kinetic energy in the laboratory system.
Same description of curves as in Fig. \ref{totsigm1}.
 }
\label{totsigm2}
\end{figure}
%%%%%%%%%%%%%%%%%%%%%%%%%%%%%%%%%%%%%%%%%%%%%%%%%%%%%%%%%%%%%

\begin{figure}[hbt]
\mbox{\epsfig{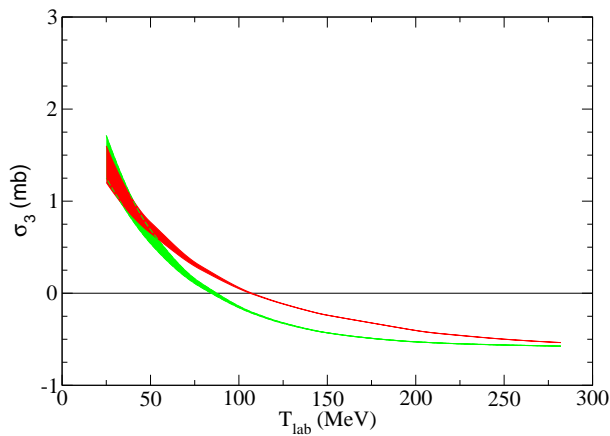}}
\caption{(Color online) Total $\bar p d$ cross section $\sigma_3$ versus the antiproton
 kinetic energy in the laboratory system.
 Results of our full calculation (including the SS+DS mechanisms)
 are shown based on the $\bar NN$ models model A (green/grey) and 
 D (red/black). The bands represent the sensitivy to variations
 of the large-angle tail of the $\bar p N$ amplitudes as discussed
 in the text. 
 }
\label{totsigm3}
\end{figure}
%%%%%%%%%%%%%%%%%%%%%%%%%%%%%%%%%%%%%%%%%%%%%%%%%%%%%%%%%%%%%
%%%%%%%%%%%%%%%%%%%%%%%%%%%%%%%%%%%%%%%%%%%%%%%%%%%%%%%%%%%%%
\begin{figure}[hbt]
\mbox{\epsfig{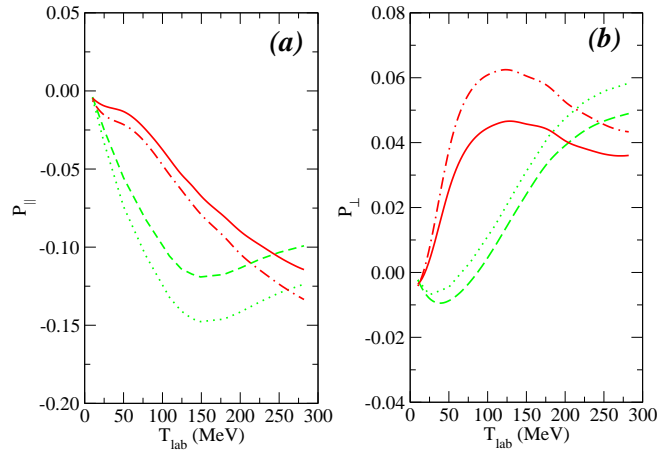}}
\caption{(Color online) Dependence of the longitudinal ($P_{||}$, panel a) and transversal ($P_{\perp}$, panel b)
 polarization on the beam energy. 
Same description of curves as in Fig. \ref{totsigm1}. 
The acceptance angle is 20 mrad.
 }
\label{fmerit}
\end{figure}
%%%%%%%%%%%%%%%%%%%%%%%%%%%%%%%%%%%%%%%%%%%%%%%%%%%%%%%%%%%%%

Results for the total unpolarized $\bar p d$ cross
section are displayed in Fig.~\ref{totsigm0} together with experimental
information \cite{BizzarriNC74,Kalogeropoulos,Burrows,Carroll,Hamilton}.
Obviously the unpolarized cross section is described 
rather well by both J\"ulich models A and D \cite{Mull2} within the SS+DS mechanisms 
(dashed and solid lines, respectively), while it is overestimated by $\sim 10-15$\% 
within the SS approximation, exemplified in Fig. \ref{totsigm0} only for model D 
(dash-dotted line).
A similar result was obtained in the spinless approximation in Ref. \cite{ujh2009}.
Taking into account the double-scatterig mechanism leads to more sizable 
changes in the results for the spin-dependent cross section,
relevant for the spin-filtering mechanism 
(see Figs.~\ref{totsigm1} and \ref{totsigm2}),
especially for $\sigma_2 $.  
One can see from Fig. \ref{totsigm2} that this cross section is reduced 
by roughly a factor of two when the double-scattering mechanism is included.
For the cross section $\sigma_1$ this difference is in the order of 10-15\%. 
Note that a decrease of the absolute values of $\sigma_0$, $\sigma_1$ and 
$\sigma_2$, due to shadowing effects, of comparable magnitude was reported in 
Ref.~\cite{salnikov} in a calculation based on the Nijmegen $\bar pN$ 
amplitudes \cite{Timmermans}. 

The cross sections $\sigma_0$, $\sigma_1$, and $\sigma_2$ 
are more or less completely determined by the $\bar pN$ at forward angles and, 
thus, can be reliably calculated within the Glauber approach. 
Indeed, the uncertainty bands turned out to be very small 
and, therefere, we don't show them in the figures. 
The tensor polarized cross section $\sigma_3$, shown in Fig.~\ref{totsigm3},
vanishes in the SS approximation. 
At low energies 25--50 MeV this cross section is in the order of 2 mb. 
Unlike the other cross sections discussed above
$\sigma_3$ turned out to be fairly sensitive to the values of the $\bar pN$ 
amplitudes at large angles, i.e. to the variations considered in sect. III A. 
Thus, there is a significant uncertainty in the predictions based on the
Glauber theory as indicated by the bands. 
With increasing energy the cross section $\sigma_3$ decreases and is only
about 0.3-0.5 mb above 100 MeV. As expected, at higher energies the sensitivity 
to the $\bar pN$ amplitudes at large angles decreases too.

With regard to the influence of the $D$-wave component of the deuteron wave
function on the total cross sections we found that its contribution
to $\sigma_1$ and $\sigma_2$ is less than 1\% 
for both considered $\bar N N$ models. The total unpolarized cross section
$\sigma_0$ decreases by $\sim 5$\% if the $D$-wave is neglected.
The cross section $\sigma_3$, which is non-zero only if the 
double-scattering mechanism is accounted for, is very sensitive to the 
$D$-wave component. 
If the $D$-wave is neglected, then $\sigma_3$ changes significantly and, 
specifically, remains positive over the whole considered energy range. 
On the other hand, we observe only minor variations when using the Paris 
deuteron wave function instead of the one of the CD-Bonn potential. 
They are smaller than the uncertainties of our predictions indicated by the 
bands in Fig.~\ref{totsigm3}.

\subsection{Polarization efficiency}

According to the analysis of the kinetics of polarization \cite{MS,NNNP1},
the polarization buildup is determined mainly by the ratio of the polarized
total cross sections to the unpolarized one \cite{MS}. Let us define the unit 
vector ${\bfg \zeta}= {\bf P}_T/ P_T$,
 where  $ {\bf P}_T={\bf P}^d$ is the  target polarization vector which 
enters Eq. (\ref{totalspin}). The non-zero antiproton beam polarization vector
${\bf P}^{\bar p}$, produced  by the polarization buildup,
 is collinear to the vector ${\bfg \zeta}$ for any directions of ${\bf P}_T$
 and can be calculated from consideration of the kinetics of polarization.
 The general solution for the kinetic
 equation for $\bar p p$ scattering is given in Ref.~\cite{MS}. Here we
 assume that this solution is valid for $\bar p d$ scattering also.
Therefore, for the spin-filtering mechanism of the polarization buildup
the polarization degree at the time $t$ is given
by \cite{MS,mss}
\begin{equation}
P_{\bar p}(t)=\tanh\left [\frac {t}{2}(\Omega_{-}^{out}-
\Omega_{+}^{out})\right ],
\label{pdeg}
\end{equation}
where
\begin{equation}
\Omega_{\pm}^{out}=nf\left \{\sigma_0\pm P_T\left [\sigma_1 +
({\bfg \zeta}\cdot {\bf \hat k})^2\sigma_2\right ]\right \}.
\label{omega}
\end{equation}

%%%%%%%%%%%%%%%%%%%%%%%%%%%%%%%%%%%%%%%%%%%%%%%%%%%%%%%%%%%%%
%%%%%%%%%%%%%%%%%%%%%%%%%%%%%%%%%%%%%%%%%%%%%%%%%%%%%%%%%%%%%
\begin{figure}[hbt]
\mbox{\epsfig{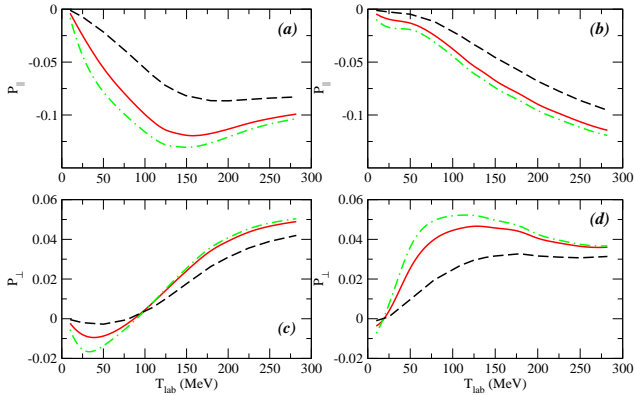}}
\caption{(Color online) Dependence of the longitudinal 
($P_{||}$, panels a and b) and transversal ($P_{\perp}$, 
panels c and d) polarization on the acceptance angle, 
for the $\bar NN$ model A (a and c) and model D (b and d). 
The dashed, solid, and dash-dotted curves are results for
acceptance angles of 10, 20, and 30 mrad, respectively.
}
\label{fmerit2}
\end{figure}
%%%%%%%%%%%%%%%%%%%%%%%%%%%%%%%%%%%%%%%%%%%%%%%%%%%%%%%%%%%%%

Here $n$ is the areal density of the target and $f$ is the beam revolving
frequency. Note that the cross sections in Eq.~(\ref{omega}) involve 
hadronic as well as Coulomb contributions, see, e.g., Refs.~\cite{MS,ujh2009}. 
Obviously the tensor cross section $\sigma_3$
from Eq. (\ref{totalspin}) does not contribute to $\Omega_{\pm}^{out}$.
Assuming the condition $|\Omega_{-}^{out}-\Omega_{+}^{out}|
<< (\Omega_{-}^{out}+\Omega_{+}^{out}$), which was found in
Refs.~\cite{MS,mss} for $\bar p p$ scattering in rings at $n=10^{14}$
cm$^{-2}$ and $f=10^6$ c$^{-1}$,
one can simplify Eq. (\ref{pdeg}).
If one denotes the number of antiprotons in the beam at the time moment $t$
as $N(t)$, then the figure of merit is $P_{\bar p}^2(t)N(t)$. This value
is maximal at the moment $t_0=2\tau$, where $\tau$ is the beam life time.
The latter is determined by $\sigma_0$, the total cross section of
the interaction of the antiprotons with the deuteron target, via  
\begin{equation}
\tau=\frac{1}{nf\sigma_0}.
\label{tau}
\end{equation}
The quantity relevant for the efficiency of the polarization buildup 
is the polarization degree $P_{\bar p}$  at the time $t_0$
\cite{mss}. In our definition for $\sigma_1$ and $\sigma_2$, which
differs from that in Refs.~\cite{MS,mss}, we find
\begin{eqnarray}
P_{\bar p} (t_0)&=&-2P_T\frac{\sigma_1}{\sigma_0}, \,\,\, {\rm if} \,\,\,
{\bfg \zeta}\cdot {\bf \hat k}=0 \ ,
\nonumber \\
P_{\bar p} (t_0)&=&-2P_T\frac{\sigma_1+\sigma_2}{\sigma_0}, \,\, {\rm if} \,\,\,
 |{\bfg \zeta}\cdot {\bf \hat k}|=1 \ . 
\label{Poldeg}
\end{eqnarray}
For evaluating the polarization degree Coulomb effects are taken into account 
via the procedure described in Ref. \cite{ujh2009}.
Thus the quantities $\sigma_i$ ($i=0,1,2$) in Eq.~(\ref{Poldeg})
are actually the sum of the hadronic cross sections, of the 
Coulomb cross section (only for $i=0$) and of the Coulomb nuclear interference terms. 
In the concrete calculation an acceptance angle
of 20 mrad is used.

The polarization degree $P_{\bar p}(t_0)$ for  ${\bfg \zeta}\cdot
{\bf \hat k}=1$ ($P_{||}$) at $P_T=P^d=1$ is shown in Fig.~\ref{fmerit} 
together with the results for ${\bfg \zeta}\cdot {\bf \hat k}=0$ ($P_\perp$).
One can see that, in general, the polarization efficiency
increases with increasing energy.
For longitudinal polarization maximal values of about 10-15\% are predicted
above 150 MeV, see Fig.~\ref{fmerit}a). 
The transversal polarization degree is smaller than the
longitudinal one for both models A and D. Of course, and as was already
pointed out in our earlier works \cite{ujh2009,jhuz2011}, there is a
significant model dependence in the predictions for both polarization cases.  

Obviously, the inclusion of the DS mechanism leads to a decrease of the 
longitudinal $P_{||}$ as well as of the transversal $P_{\perp}$ polarization 
efficiences by about 20-30\% as compared to the SS mechanism alone \cite{ujhprm}.
Nevertheless, for both considered models
the magnitude of the spin-dependent cross sections is still comparable or 
even larger than those for $\bar p p$ \cite{mss,ujh2009,jhuz2011}.
In this context let us also mention that 
the values for the polarization degree we obtained are somewhat smaller
than those presented in \cite{salnikov}, based on the Nijmegen $\bar NN$ 
partial wave analysis \cite{Timmermans} from 1994.

Finally, in Fig.~\ref{fmerit2} we document the dependence of the quantities
$P_{||}$ and $P_{\perp}$ on the acceptance angle. As expected, in general
the polarization degree increases with increasing acceptance angle. 
But the variations themselves are not too dramatic. 

\section{Summary}

 In the present work we analyzed the role of the spin dependence of 
 the $\bar pN$ amplitude in elastic $\bar p d$ scattering for energies of 
 50--300 MeV of the incident antiproton on the basis of the Glauber 
 theory. In the actual calculations we utilized elementary $\bar pN$ 
 amplitudes generated from the J\"ulich $\bar NN$ model \cite{Mull2}.
 The $S$- and $D$-wave components of
 the deuteron were included into the calculation and the single- and 
 double-scattering mechanisms were taken into account. 
 
Since some of the spin-dependent amplitudes exhibit a non-diffractive behaviour
we performed various test calculations in order to pin down the angular range 
where the Glauber theory can be reliably applied.
Thereby, it turned out that this approach works rather well for the 
region of the forward peak, over the whole considered energy region.
Obviously,
for the considered ${\bar p} N$ models those amplitudes with non-diffractive
character are fairly small as compared to the dominant spin-independent amplitude 
(with a pronounced diffractive behaviour) so that the former do not
spoil the applicability of the Glauber theory.
This means, in turn, that the approach can be used savely for the calculation 
of the total spin-dependent $\bar p d$ cross sections via the optical theorem. 
 With regard to the considered $\bar p d$ differential cross sections 
and vector- and tensor analyzing powers ($A_y^d$, $A_y^{\bar p}$, $A_{yy}$, $A_{xx}$)
our investigation indicates that reliable predictions can be obtained for 
c.m. scattering angles up to 
$50^\circ-60^\circ$ in the $\bar p d$ system. For 179~MeV, where
data on the differential cross section exist, this range covers 
the first minimum and the onset of the second maximum. Here our
results based on the $\bar p N$ amplitudes of the J\"ulich model D 
turned out to agree nicely with the experiment while model A 
overestimates the measured $\bar p d$ cross section at the minimum. 

The total polarized $\bar p d$ cross sections $\sigma_i$ ($i=1,2,3$), and
specifically, the polarization degree of the antiproton beam is of 
interest in the context of plans to establish a polarized antiproton beam
via the spin-filtering method as proposed by the PAX collaboration 
(see also \cite{Wojt}). 
Corresponding predictions presented in this work, exhibit a sizable model 
dependence, reflecting the uncertainties in the spin dependence of the 
elementary $\bar p p$ and $\bar p n$ interactions. Still, for both considered 
models we find that the magnitude of the spin-dependent cross sections is
comparable or even larger than those for $\bar p p$.
Thus, our results suggest that deuteron targets can be used for
the polarization buildup of antiprotons at beam energies of 100--300 MeV
with similar and possibly even higher efficiency than $\bar p p$ scattering.
Nonetheless, only concrete experimental data on the spin-dependent part of the 
cross sections of $\bar pp $ and $\bar p d$ scattering
will allow one to confirm or disprove the feasibility of the spin
filtering mechanism for the antiproton polarization buildup.

\section{Acknowledgements}
 We acknowledge stimulating discussions with N.N.~Nikolaev and F.~Rathmann.
 We would like to thank V.I.~Kukulin and M.N.~Platonova for 
 communicating more details concerning 
 the work published in \cite{pkuk,pkuk2}.
 This work was supported in part by the WTZ project no. 01DJ12057. 

\section*{Appendix: Invariant amplitudes $A_i$ and spin observables 
for $\bar p d$ elastic scattering}
\setcounter{equation}{0}
\renewcommand{\theequation}{A.\arabic{equation}}
 
The scattering matrix ${M}_{\bar pd}$ in Eq.~(\ref{glafi})
can be expressed via the 12 invariant amplitudes $A_i$ in the following way \cite{pkuk}:
\begin{eqnarray}
\label{Ai}
{M}_{\bar p d}&=&(A_1+A_2{\bfg\sigma}\cdot {\bf \hat n})+
(A_3+A_4\bfg\sigma \cdot {\bf \hat n})({\bf S}\cdot {\bf \hat q})^2 \nonumber \\
&+&(A_5+A_6\bfg\sigma \cdot {\bf \hat n})({\bf S}\cdot {\bf \hat n})^2+
(A_7\bfg\sigma \cdot {\bf \hat k})({\bf S}\cdot {\bf \hat k})\nonumber \\
&+&(A_8\bfg\sigma \cdot {\bf \hat q})[({\bf S}\cdot {\bf \hat q})({\bf S}\cdot {\bf \hat n})
+({\bf S}\cdot {\bf \hat n})({\bf S}\cdot {\bf \hat q})]\nonumber \\
&+&(A_9+A_{10}\bfg\sigma \cdot {\bf \hat n})({\bf S}\cdot {\bf \hat n})
+(A_{11}\bfg\sigma \cdot {\bf \hat q})({\bf S}\cdot {\bf \hat q})\nonumber \\
&+&(A_{12}\bfg\sigma \cdot {\bf \hat k})[({\bf S}\cdot {\bf \hat k})({\bf S}\cdot {\bf \hat n})
+({\bf S}\cdot {\bf \hat n})({\bf S}\cdot {\bf \hat k})] \ . \nonumber \\
\end{eqnarray}
Here
${\bf S}=(\bfg \sigma_p+\bfg \sigma_n)/2$ is the total spin of the deuteron
and the definition of the unit vectors ${\bf \hat k}$, ${\bf \hat q}$, and ${\bf \hat  n}$
is given right after Eq.~(\ref{pbarN}).
In the coordinate system as chosen in Ref.~\cite{pkuk}, with the axes
${\bf\hat e_x}={\bf \hat q}$, ${\bf \hat e_y}={\bf \hat n}$, and 
${\bf \hat e_z}={\bf \hat k}$ the differential cross section $d\sigma/dt$
and analyzing powers $A_y^{\bar p}$, $A_y^d$, $A_{xx}$, and $A_{yy}$ 
take the following forms \cite{pkuk}
\begin{eqnarray}
\label{diffcs}
d\sigma/dt&\equiv& \frac{1}{3}\Sigma\nonumber \\
&=&|A_1|^2+|A_2|^2+\frac{2}{3}\Bigl \{Z+Re[2A_1^*(A_3+A_5)\nonumber \\
&+&2A_2^*(A_4+A_6)+A_3^*A_5+ A_4^*A_6]\Bigr \},
\end{eqnarray}
where $Z=\sum_{i=3}^{12}|A_i|^2$.
\begin{eqnarray}
\label{ayp}
A_y^{\bar p}&=&2Re[2(A_1^*+A_3^*+A_5^*)(A_2+A_4+A_6)\nonumber \\
&+&A_1^*A_2-A_3^*A_6-A_4^*A_5+2A_9^*A_{10}]\Sigma^{-1},
\end{eqnarray}
\begin{eqnarray}
\label{ayd}
A_y^d&=&2Re[(2A_1^*+A_3^*+2A_5^*)A_9 +(2A_2^*+A_4^* \nonumber \\
&+&2A_6^*)A_{10} +A_7^*A_{12}+A_8^*A_{11}]\Sigma^{-1},
\end{eqnarray}

\begin{eqnarray}
\label{ayy}
A_{yy}&=&\{2(|A_5|^2+|A_6|^2 + |A_9|^2+|A_{10}|^2) \nonumber \\
&-&(|A_3|^2+|A_4|^2 + |A_7|^2 + |A_8|^2\nonumber \\
&+&|A_{11}|^2+|A_{12}|^2)+\nonumber \\
&+&2Re[A_1^*(2A_5-A_3)+A_2^*(2A_6-A_4)\nonumber \\
&+&A_3^*A_5+A_4^*A_6]\}\Sigma^{-1}, 
\end{eqnarray}
\begin{eqnarray}
A_{xx}&=&\{2(|A_3|^2+|A_4|^2 + |A_{11}|^2+|A_{12}|^2)\nonumber \\
&-&(|A_5|^2+|A_6|^2 + |A_7|^2 + |A_8|^2\nonumber \\
&+&|A_{9}|^2+|A_{10}|^2)+\nonumber \\
&+&2Re[A_1^*(2A_3-A_5)+A_2^*(2A_4-A_6)\nonumber \\
&+&A_3^*A_5+A_4^*A_6]\}\Sigma^{-1}. 
\label{axx}
\end{eqnarray}

The amplitudes $A_i$ can be rewritten as linear combinations of the amplitudes $F_i$ 
used in our previous paper \cite{ujh2009}. The latter are defined within 
a different basis as compared to Eq.~(\ref{Ai}).
In terms of the amplitudes $F_i$ the observables
given in Eqs.~(\ref{diffcs})--(\ref{axx}) coincide with those given in the Appendix of
Ref.~\cite{ujh2009}.
For collinear kinematics ($q=0$) the scattering matrix $M_{\bar p d}(0)$
contains four independent terms \cite{rekalo} and can be written as \cite{uzepan98} 
\begin{eqnarray}
\label{fab}
{M}_{\bar p d; \alpha\beta}(0)=g_1\delta_{\alpha\beta}+
(g_2-g_1) \hat k_\alpha \hat k_\beta+ \nonumber \\
ig_3{\sigma}_i\epsilon_{\alpha\beta  i}+ i(g_4-g_3){\sigma}_i \hat k_i \hat k_j
\epsilon_{\alpha\beta  j},
\end{eqnarray}
where $ \sigma_i$ ($i=x,y,z$) are the Pauli spin matrices acting on 
the spin states of the antiproton,
$\epsilon_{\alpha \beta \gamma}$ is the fully antisymmetric tensor,
$\hat k_\alpha$ are the Cartesian components of a unit vector ${\bf {\hat k}}$
pointing along the beam momentum, and $g_i$ $(i=1,\dots,4)$ are complex amplitudes.
When considering  Eqs. (\ref{Ai}) and (\ref{fab}) together, one can find that 
in collinear kinematics with $OZ||{\bf \hat k}$:
$A_2=A_4=A_6=A_8=A_9=A_{12}=0$, $A_3=A_5$, $A_{10}=A_{11}$. The independent 
amplitudes $g_i$ are related with the amplitudes $A_i$ via 
\begin{eqnarray}
\label {gi}
g_1&=&A_1+A_3, \,\, g_2=A_1+A_3+A_5,\nonumber \\
 g_3&=&-A_{10},\, \, g_4=-A_7 \ .
\end{eqnarray}
%%%
Taking into account the proper normalization of the scattering matrix $M_{\bar p d}$ 
and using those relations between the $A_i$'s and $g_i$'s,
the total hadronic cross sections in Eq.~(\ref{totalspin}) given in Ref.~\cite{ujh2009}
on the basis of the generalized optical theorem 
can be rewritten as
\begin{eqnarray}
\label {sigmai}
\sigma_0&=&\frac{4}{3}\sqrt{\pi} \, Im(2g_1+g_2) \ ,\nonumber \\
\sigma_1&=&-4\sqrt{\pi} \,  Im(g_3) \ ,\nonumber \\
\sigma_2&=&-4\sqrt{\pi} \, Im(g_4-g_3) \ ,\nonumber \\
\sigma_3&=&4\sqrt{\pi} \, Im(g_1-g_2) \ .
\end{eqnarray}
Please note that the expressions for the $\sigma_i$ presented in Eqs.~(19)-(20)
of our previous work [8] are erroneous. Specifically, the
correct signs of $\sigma_1$ and $\sigma_2$ are opposite to
those given in Ref. [8].
Numerically those errors have practically no influence on the value of $\sigma_0$
and also not on the absolute values of the polarization efficiencies
in the considered energy region of 50-300 MeV, as we verified in corresponding
computations.

%%%%%%%%%%%%%%%%%%%%%%%%%%%%%%%%%%%%%%%%%%%%%%%%%%%%%%%%%%%%%

\end{document}